\documentclass[sigconf]{acmart}

\makeatletter
\def\@ACM@checkaffil{
    \if@ACM@instpresent\else
        \ClassWarningNoLine{\@classname}{No institution present for an affiliation}%
    \fi
    \if@ACM@citypresent\else
        \ClassWarningNoLine{\@classname}{No city present for an affiliation}%
    \fi
    \if@ACM@countrypresent\else
        \ClassWarningNoLine{\@classname}{No country present for an affiliation}%
    \fi
}
\makeatother

\usepackage{xspace}
\usepackage{multirow}
\usepackage{makecell}
\usepackage{framed}
\usepackage{tabularx} 
\usepackage{subcaption} 
\usepackage[utf8]{inputenc}
\usepackage[russian,english]{babel}

\newcommand\eg{\emph{e.g.},\xspace}
\newcommand\ie{\emph{i.e.},\xspace}
\providecommand{\etal}{\emph{et al.}\xspace}

\AtBeginEnvironment{quote}{\par\small\itshape}

\newcommand{\takeaway}[1]{\begin{framed}\noindent{\textbf{Takeaway.} #1}\end{framed}}

\newcommand{\taDataStart}{May 12, 2021\xspace}
\newcommand{\taDataEnd}{Sep 20, 2022\xspace}
\newcommand{\taCrawlStart}{Mar 12, 2022\xspace}
\newcommand{\taCrawlEnd}{Sep 30, 2022\xspace}

\newcommand{\taTotal}{7,330\xspace}
\newcommand{\taThreads}{1,319\xspace}
\newcommand{\taUsers}{1,229\xspace}
\definecolor{taColor}{HTML}{27a577}
\newcommand{\gmapsDataStart}{Jan  1, 2020\xspace}
\newcommand{\gmapsDataEnd}{Jun 30, 2022\xspace}
\newcommand{\gmapsCrawlStart}{Mar  4, 2022\xspace}
\newcommand{\gmapsCrawlEnd}{Jun 30, 2022\xspace}

\newcommand{\gmapsTotal}{2,200,368\xspace}
\newcommand{\gmapsPlaces}{122,826\xspace}
\newcommand{\gmapsUsers}{1,164,002\xspace}
\newcommand{\gmapsTowns}{8,660\xspace}
\definecolor{gmapsColor}{HTML}{ea4335}
\newcommand{\vkDataStart}{Mar 18, 2022\xspace}
\newcommand{\vkDataEnd}{Jun 30, 2022\xspace}
\newcommand{\vkCrawlStart}{Mar 18, 2022\xspace}
\newcommand{\vkCrawlEnd}{Jun 30, 2022\xspace}

\newcommand{\vkTotal}{3,963,981\xspace}
\newcommand{\vkUsers}{919,172\xspace}
\newcommand{\numManualLabelled}{3,200\xspace}
\newcommand{\numManualLabelledTA}{660\xspace}
\newcommand{\numManualLabelledGmaps}{2,540\xspace}
\definecolor{vkColor}{HTML}{0073f7}

\usepackage{graphics}

\AtBeginDocument{%
  \providecommand\BibTeX{{%
    \normalfont B\kern-0.5em{\scshape i\kern-0.25em b}\kern-0.8em\TeX}}}

\setcopyright{acmcopyright}
\copyrightyear{2018}
\acmYear{2018}
\acmDOI{XXXXXXX.XXXXXXX}

\acmConference[Conference acronym 'XX]{Make sure to enter the correct
  conference title from your rights confirmation email}{June 03--05,
  2018}{Woodstock, NY}
\acmPrice{15.00}
\acmISBN{978-1-4503-XXXX-X/18/06}

\usepackage{enumitem}
\begin{document}

\title{Reviewing War: Unconventional User Reviews as a Side Channel to Circumvent Information Controls}

\author{José Miguel Moreno}
\email{josemore@pa.uc3m.es}
\affiliation{%
  \institution{Universidad Carlos III de Madrid}
}

\author{Sergio Pastrana}
\email{spastran@inf.uc3m.es}
\orcid{0000-0003-1036-6359}
\affiliation{%
  \institution{Universidad Carlos III de Madrid}
}

\author{Jens Helge Reelfs}
\email{reelfs@b-tu.de}
\orcid{0000-0003-4635-8561}
\affiliation{%
  \institution{Brandenburg University of Technology (BTU) Cottbus}
}

\author{Pelayo Vallina}
\email{pelayo.vallina@imdea.org}
\orcid{0000-0001-7551-7176}
\affiliation{%
  \institution{IMDEA Networks Institute /\\ Universidad Carlos III de Madrid}
}

\author{Andriy Panchenko}
\email{andriy.panchenko@b-tu.de}
\affiliation{%
  \institution{Brandenburg University of Technology (BTU) Cottbus}
}

\author{Georgios Smaragdakis}
\email{g.smaragdakis@tudelft.nl}
\orcid{0000-0002-4127-3617}
\affiliation{%
  \institution{TU Delft}
}

\author{Oliver Hohlfeld}
\email{oliver.hohlfeld@b-tu.de}
\orcid{0000-0002-7557-1081}
\affiliation{%
  \institution{Brandenburg University of Technology (BTU) Cottbus}
}

\author{Narseo Vallina-Rodriguez}
\email{narseo.vallina@imdea.org}
\orcid{0000-0002-5420-6835}
\affiliation{%
  \institution{IMDEA Networks Institute}
}

\author{Juan Tapiador}
\email{jestevez@inf.uc3m.es}
\orcid{0000-0002-4573-3967}
\affiliation{%
  \institution{Universidad Carlos III de Madrid}
}


\setcopyright{none}
\settopmatter{printacmref=false, printccs=false, printfolios=false}
\renewcommand\footnotetextcopyrightpermission[1]{}
\pagestyle{plain}
\acmConference{}{}{}
\renewcommand{\shortauthors}{}

\begin{abstract}
During the first days of the 2022 Russian invasion of Ukraine, Russia's media regulator blocked access to many  global social media platforms and news sites, including Twitter, Facebook, and the BBC. To bypass the information controls set by Russian authorities, pro-Ukrainian groups explored unconventional ways to reach out to the Russian population, such as posting war-related content in the user reviews of Russian business available on Google Maps or Tripadvisor. This paper provides a first analysis of this new phenomenon by analyzing the creative strategies to avoid state censorship. Specifically, we analyze reviews posted on these platforms from the beginning of the conflict to September 2022. 
We measure the channeling of war messages through user reviews in Tripadvisor and Google Maps, as well as in VK, a popular Russian social network. Our analysis of the content posted on these services reveals that users leveraged these platforms to seek and exchange humanitarian and travel advice, but also to disseminate disinformation and polarized messages. Finally, we analyze the response of platforms in terms of content moderation and their impact. 
\end{abstract}




\maketitle

\section{Introduction}
\label{sec:introduction}

With the beginning of the full-scale invasion of Ukraine on 24 February 2022, Roskomnadzor --- Russia's media regulator --- implemented information control measures to block social media platforms like Facebook and Twitter, and news websites like the BBC~\cite{WiredRussiaBlockSocialMedia}. These measures were seen not only as an attempt to stop the dissemination within Russia of any information not provided by official sources, but also as retaliation for the removal of Twitter and Facebook accounts allegedly belonging to two pro-Russian disinformation groups~\cite{RussianTrollFarm,NBCNewsAccountRemoval}, and EU bans on Russia's news outlets Russia Today and Sputnik~\cite{RTSputnikBan}.
The Open Observatory of Network Interference (OONI) confirmed the deployment of censorship mechanisms by Russian Internet Service Providers (ISPs) from the beginning of the 2022 conflict in a report dated March 7, 2022~\cite{OONIReport}.

Information controls are frequent in time of war, and so are evasive manoeuvres to bypass them. However, Russia's censorship efforts were answered with some inventive proposals. On February 28, an account presumably affiliated with \texttt{Anonymous} suggested to employ user reviews in restaurants and other business located via Google Maps to deliver war-related information to the Russian population~\cite{AnonCallForReviews}. Tinder, Tripadvisor, and Telegram were also targeted as means to reaching out to the Russian population, thus bypassing the strict media control implemented by Russian authorities~\cite{WiredRussianPropagandaWall}. On March 4, the \texttt{squad303} group offered the possibility to target millions of Russian citizens with SMS via the \url{1920.in} site. This service was later extended to send emails and Whatsapp or Viber messages.
Some prominent online service providers responded to these campaigns by actively removing war-related content from their platforms. Google and Tripadvisor placed restrictions on reviews of Russian business, and Google Maps soon stopped accepting new reviews for places located in Russia, Ukraine, and Belarus. They argued that such reviews violate company policies~\cite{GoogleTripadvisorDisableContent,tripadvisorCEOletter,wsjTripadvisorGoogleMaps}.

The creative use of online services as side-channels to bypass Russian information controls on the web was anecdotally echoed by the media. Yet, there is no quantitative or qualitative assessment of the user involvement, intentions, and intensity of  these campaigns, nor the response by platform operators to moderate content. In this paper, we answer these questions by crawling and analyzing several datasets purposely collected from Google Maps, Tripadvisor, and the biggest Russian social network, Vkontakte (VK). We collect data from February through September, 2022, and we also collect historical data for comparison (\S\ref{sec:datasets}).
Our key contributions and findings include:
\begin{itemize}[leftmargin=*]
    \item We study changes in the volume of reviews and moderated content in Google Maps and Tripadvisor (\S\ref{sec:volume}). Our analysis indicates an increment of nearly 100$\times$ in new posts for Tripadvisor during the early days of the war.
    \item We leverage text-based analysis techniques to label reviews as related to war or not (\S\ref{subsec:war-related}). We find evidence that content posted in these platforms is used to deliver political messages related to the war (\ie to counter misinformation/propaganda) to Russian audiences. Topic analysis of posted messages (\S\ref{subsec:topic-analysis}) confirms a noticeable change in user discourse in Tripadvisor and Google Maps. This ratifies the use of these platforms as channels to disseminate war-related content.
    \item We conduct a qualitative assessment of the content (\S\ref{sec:case-studies}) and find that messages can be grouped in four main categories: (dis)information campaigns, humanitarian help, travel advice, and polarized/hate speech. We leverage network analysis technique to find evidence of organized campaigns to disseminate slogans (\S\ref{sec:org_campaigns}).
    \item Finally, we study the reaction of service operators to control or
remove war-related information (\S\ref{sec:moderation}). We find that Tripadvisor
monitors and, if needed, removes war-related content not
abiding by their Terms of Service (\eg hate speech) from their
forums in less than two days on average. Also, we observe that
the number of reviews in Google Maps was reduced up to
an average of 8 reviews per day in conflict areas. 
\end{itemize}

\section{Background and Related Work}
\label{sec:timeline}

State censorship during times of military conflicts and in oppressive regimes has been subject of attention and analysis for a long time~\cite{price1942governmental,morgans2017freedom,pearce2017global,marczak2015analysis,niaki2020iclab}. The 2014 Russia-Ukraine conflict offered a valuable case study of Russia's information war strategy. Russia started offensive cyber-operations against Ukraine not later than 2009 as a part of a broader war campaign against NATO and EU countries~\cite{unwala2016brandishing}. In 2014, information war operations intensified against Ukraine~\cite{volkova2016account}.

While initially aiming at spreading misinformation and propaganda, with a start of a full-scale attack on Ukraine, Russian authorities boosted media control by blocking free access to the Internet. According to OONI's web connectivity public data~\cite{ooni},
sites such as \texttt{bbc.com} and \texttt{facebook.com} were blocked on the 4th of March 2022, while others such as \texttt{twitter.com} were censored just two days after the beginning of the armed conflict. Independent Russian news channels like \texttt{currenttime.tv} or \texttt{tvrain.ru}, censorship evasion tools and VPN services were also blocked in late February 2022.
Yet, popular Russian social networks like 
VKontakte (VK) and western ones such as Instagram and 
YouTube remained accessible to limit collateral damage. Some of them were gradually blocked 
in Russian ISPs as the conflict evolved~\cite{nytRussiaMediaCensorship}.
This was done in fear of civil protests against the war, preceded by repression and mass arrests on March 4 2022~\cite{bbcWarProtestsInRussia}. 
In fact, words such as \textit{war} and \textit{invasion} were officially banned in Russia's media.   

The intensification of information controls by Russian authorities 
motivated the creative and novel use of non-blocked side-channels.
A campaign lead by the group \textit{Anonymous} on Twitter provided information and called for these actions in Google Maps~\cite{AnonCallForReviews}, providing possible messages to spread over Russian places. By October '22, this tweet has been retwetted more than 27k times, with more than 79k likes. As a result, these websites soon became a niche for spreading information about the war. 
In fact, on March 2, 2022, Tripadvisor's CEO recommended the use of Ukraine and Russia forums to \textit{``enable users to share information''} about the situation in the country~\cite{tripadvisorCEOletter}. However, Tripadvisor staff soon posted messages for certain Russian places indicating that reviews were disabled due to high volume of war-related content, and that users should use the forums to inform about available travel options within Ukraine instead~\cite{wsjTripadvisorGoogleMaps}.
Since then, the travel forums for Russia and Ukraine turned into a platform for discussion about the war situation.

\noindent{\bf Related work:} Recent research efforts have studied the effects of the 2022 Russia-Ukraine conflict from different angles.
Xue \etal developed a novel approach to detect and measure censorship on RuNet enforced by the Russian Government at different network levels~\cite{TSPU-IMC2022}.
Jonker \etal focused on the infrastructure powering websites under Russian ccTLDs (\eg ``.ru'') and how the invasion of Ukraine has triggered a \textit{repatriation} of such infrastructure towards national providers \cite{Whereru}.
In the cybercrime field, Vu \etal investigated the role of criminal groups in this conflict, finding that they actively engaged in hacktivism during its initial phase and then lost interest~\cite{Bored-Cyberwar}.
Finally, Hanley\etal studied the influence of Russian propaganda websites on political subreddits during the conflict~\cite{Happenstance}. Our analysis complements these efforts and helps to draw a richer picture of how online services were used during the war and how they were used to avoid state-level information controls.

\vspace{-1mm}
\section{Datasets}
\label{sec:datasets}

Due to anecdotal evidence suggesting that both Google Maps and Tripadvisor were being used to disseminate war-related information~\cite{wsjTripadvisorGoogleMaps,tripadvisorCEOletter}, we focus our study on these platforms. We note that both services were (and still are, as of October'22) available to Russian citizens. 
We also decided to monitor the activities in VK, the largest Russian social network, in order to better understand potential social activities around the war in a Russian site and compare trends. 
This section describes our crawling efforts and provides statistics of the dataset obtained.

\vspace{1mm}
\noindent\textbf{Tripadvisor.}
We crawl Ukraine and Russian travel forums in Tripadvisor for seven months (from \taCrawlStart to \taCrawlEnd), split in two periods. First, we conduct regular crawls for a period of two months (from \taCrawlStart to May 12, 2022) harvesting all the posts made since the beginning of war. As some posts were being removed by Tripadvisor due to infringement of its Terms of Service, the crawler checked for new posts every hour and collected any new content, thus allowing us to flag and measure removals. This crawling period allowed us to conduct online monitoring on the platform.
We also conducted one single crawl to obtain pre-war posts since May 12, 2021. Due to technical limitations and a decrease in post volume over time (\S\ref{subsec:crawl-limitations}), we decided to stop the data collection, and resume it in September 2022 with a lower crawling frequency. Overall, the dataset contains \taTotal posts made in \taThreads different threads by \taUsers different users.

\vspace{1mm}
\noindent\textbf{Google Maps.}
Using a purpose-built Chromium-based crawler, we harvested \gmapsTotal reviews obtained from \gmapsPlaces locations in Russia. We started crawling these reviews on \gmapsCrawlStart{}.
We fetch new reviews every 2 hours and update the list of places daily.
Our Chromium-based instrumentation makes use of the ``Nearby'' search functionality from Google Maps to  lists any places (\eg hotels, restaurants, museums) found in a given town or its vicinity. 
We feed the crawler with a seed formed by 321 predefined Russian towns~\cite{Russian-towns}
from where to discover places. In the end, combining these two methods we
covered \gmapsTowns different towns. There are \gmapsUsers unique users with at least one posted review in the dataset. We stopped crawling on Jun 30 because the activity on this platform had stopped
as we mention in Section~\ref{sec:introduction}.

\vspace{1mm}
\noindent\textbf{VK.}
We crawled \vkTotal posts appearing on the top-50 public VK communities with the most followers at a given date. It contains publications and replies from 51 different communities, published by \vkUsers different users. We started crawling this social network on \vkCrawlStart{}.
Given the intensity of these communities, we follow a best-effort approach to fetch new posts every 15 minutes, updating the list of communities daily.
We acknowledge that, in some posts with high activity, our crawler may miss some replies as VK limits the number of most recent messages that can be accessed.

\begin{table}[t]
    \centering
    \caption{Datasets with their respective volume, crawling and data periods.}
        \begin{tabular}{lllr}
            \toprule
            \textbf{Dataset} &
            \makecell[l]{\textbf{Crawling}\\\textbf{period}} &
            \textbf{Data period} &
            \textbf{Size} \\
            \midrule

            Tripadvisor &
            \makecell[l]{\taCrawlStart{} -\\ \taCrawlEnd} &
            \makecell[l]{\taDataStart{} -\\ \taDataEnd} &
            \makecell[r]{\taTotal\\posts}\\

            Google Maps &
            \makecell[l]{\gmapsCrawlStart{} -\\ \gmapsCrawlEnd} &
            \makecell[l]{\gmapsDataStart{} -\\ \gmapsDataEnd} &
            \makecell[r]{\gmapsTotal\\reviews}\\

            VK &
            \makecell[l]{\vkCrawlStart{} -\\ \vkCrawlEnd} &
            \makecell[l]{\vkDataStart{} -\\ \vkDataEnd} &
            \makecell[r]{\vkTotal\\posts}\\

            \bottomrule
        \end{tabular}%
    \label{tab:dataset}
\end{table}

\subsection{Limitations}
\label{subsec:crawl-limitations}

Despite our best-effort to collect data, we faced some technical challenges during our crawling.
Google Maps soon disallowed reviews in Russian places
at the beginning of March
due to the high amount of off-topic messages~\cite{wsjTripadvisorGoogleMaps}. This caused a drop in the volume of traffic. Therefore, we cannot reliably determine which fraction of new messages would be focused on war-related content should the moderation policy had not been implemented.
In the case of Tripadvisor, we were able to collect data every hour for the first two months. At the beginning of May we experienced temporary bans on our crawler, which we did not attempt to circumvent due to ethical reasons (\S\ref{sec:ethics}). We extended the frequency of crawls to one per day. Afterwards, we observed a decrease in the number of war-related posts in June (see \S\ref{sec:contentanalysis}), and decided to stop the crawling and resume it in September to check again. Despite these interferences in the crawling period, we are able to collect historical data written by forum users, and thus this does not affect most of our analyses. We cannot, however, measure removals with precision as we discuss in \S\ref{sec:moderation}.
For VK, our crawler gets messages for each of the top-50 (by number of followers) public communities. Due to VK restricting the access only to the most recent messages, we cannot know if a missing message is not shown due to it being removed, or just because it is not recent.
Finally, while VK provides the number of impressions and likes for each message, Tripadvisor and Google Maps do not provide equivalent figures, which prevents us to reliably measure the impact of each message.

\subsection{Ethical Considerations}
\label{sec:ethics}
Data was collected by automatic means (crawlers) using standard ethical guidelines~\cite{kenneally2012menlo}. We used sequential crawlers and did not produce more traffic to the servers as a human user would do. 
The dataset used in this study might contain sensitive data, including Personally Identifiable Information (PII) such as usernames.
Unfortunately, it is impossible to obtain informed consent from all users. According to the British Society of Criminology Statement on Ethics, we do not require informed consent from the participants because the dataset $(i)$ is publicly accessible; and $(ii)$ will be used for research on collective behavior, without aiming to identify particular members. However, as this study involves analyzing content generated by human subjects, it does require ethical review. We applied and obtained approval from our IRB using the following research protocol:
(1) The data is stored in our secured private servers with strict access control mechanisms.
(2) We do not store users' personal identifiers, which are replaced by a unique hash value obtained from their usernames. No efforts to deanonymize the data are carried out.
(3) We do not store or visualize any images or other multimedia material posted on the crawled platforms.
(4) The data will be used only for the purpose of this research.

The research offers benefits in terms of facilitating a better understanding of techniques to bypass censorship and other information controls. Our observations will positively inform our current understanding of this topic.

\section{Content analysis}\label{sec:contentAnalysis}
\label{sec:contentanalysis}

In this section we present the analysis of the three datasets. Our analysis focuses on four aspects: $(i)$ the observed traffic shifts that correlate with the beginning of the war (\S\ref{sec:volume}); $(ii)$ the analysis of how much new content is related to the war (\S\ref{subsec:war-related}); $(iii)$ the topics being discussed (\S\ref{subsec:topic-analysis}); and $(iv)$ the Russian places that are targeted Google Maps with war-related content (\S\ref{subsec:geography}).

\vspace{-2mm}
\subsection{Activity Analysis}
\label{sec:volume}

\begin{figure}[t!]
    \centering
    \includegraphics[width=\columnwidth]{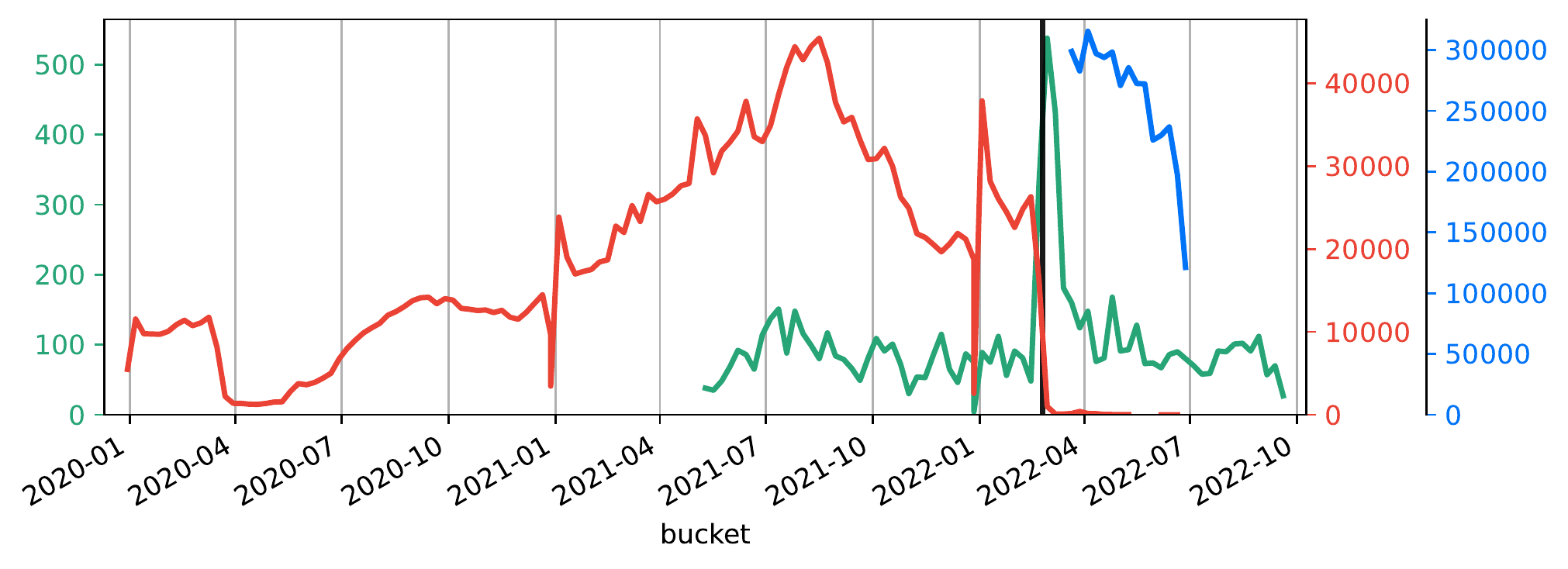}
    \vspace*{-2em}
    \caption{
        Volume of Tripadvisor posts (in \textcolor{taColor}{green}), Google Maps reviews (in \textcolor{gmapsColor}{red}) and VK posts (in \textcolor{vkColor}{blue}). The vertical
        bold line represents the beginning of the invasion.
    }
    \label{fig:volume-all}
\end{figure}

Figure~\ref{fig:volume-all} shows
the number of posts or reviews that were published in each of the three studied platforms.
For both Tripadvisor and Google Maps, we see a clear change in volume right after the beginning of the war on Feb 24 (vertical black line in Figure~\ref{fig:volume-all}), this change manifests differently across platforms in terms of intensity.
In the case of Google Maps, there is a clear drop in the daily amount of published reviews, suggesting some kind of content moderation, which is consistent with Google's policy of not accepting new reviews for places located in Russia, Ukraine, and Belarus. 
For Tripadvisor, instead, we see a slight increase in the number of posts due to a larger volume of war-related posts in these travel forums, which was indeed allowed by forum administrators (See \S\ref{subsec:war-related}).
In the case of VK, as we lack historical data, we cannot determine if the war triggered a change in VK's traffic volume. We observe, however, a significant drop of posts by the end of June 2022, when we decided to stop the crawler.

\subsection{Labeling War-related Content}\label{subsec:war-related}

We analyzed the content of the posted messages to determine how much of the observed traffic volume increase can be attributed to war-related content.
To do so, we apply a combined approach using both quantitative and qualitative methods. We start with a manual analysis of a subset of messages (70\% threads of conversations from Tripadvisor and 80\% of reviews from Google Maps) posted during the first 2 months of the conflict. This allows us to get an overview of the most common topics being discussed, which we describe in detail in Section~\ref{sec:case-studies}.

Though informative, such time-consuming manual labelling process does not scale. 
To overcome this challenge, we first employed unsupervised topic mining techniques~\cite{grootendorst2022bertopic} (masked LM embedding, dimension reduction, clustering), which indeed identified topics related to war. However, the results were very specific and contained huge amounts of false positives (FP).
Instead, we opted for an automated labelling approach based on a set of keywords obtained from our qualitative evaluation. In this process we aim at reducing the amount of FP---\ie \emph{better be safe than sorry}.

The methodology to build the classifier is as follows: First, we took a random sample of \numManualLabelled{} messages (\numManualLabelledTA{} from Tripadvisor and \numManualLabelledGmaps{} from Google Maps) and manually label them as `war' or `non-war' related. We consider a message is war-related if it clearly and explicitly discusses aspects of the conflict, \eg giving advice to refugees, positioning themselves pro or against the invasion, or trying to bypass Russian censorship.
Second, we normalize all messages and we extract the most common keywords present in war-related messages. Our normalization pipeline consists on removing URLs, punctuation symbols, and any character that is not used in either the English, Cyrillic or Polish alphabets.
Third, we score each keyword according to the following criteria:
$(i)$ 3 points to war-like keywords---up to a maximum of 6 points,
$(ii)$ 2 points to violence-related keywords---up to 4 points, and
$(iii)$ 1 point to keywords that, while are frequently occurring, do not belong to any of the previous---up to 2 points.
We cap the score per group of keywords to prevent long messages from gaining an unfair advantage. We also slightly increase the final score of messages with very few keywords (\eg \emph{``Glory to Ukraine''}).
Additionally, we use ``correct keywords'' with negative weights to minimize the number of FP.
The list of keywords and weights (\ie scores) given to each of them is listed in Appendix~\ref{appendix:keywords}.
We get a score for each message based on the number of keywords found and their scores. We empirically determined a classification threshold of 5, meaning that we classify a message as war related if it has a score $\geq 5$, and non-war related otherwise. This threshold was fine-tuned after several iterations conducting manual validation on the posts. 

We measure classification quality via a set of lightweight crowdsourced campaigns.
Due to the predominant presence of Russian language on both Google Maps and VK posts, we employ \emph{(i)} non-native speaking coders using machine translation, and \emph{(ii)} verify the results with a second labelling pass from native-speaking expert coders.
The campaigns were set up for each dataset by carefully sampling sets of each 25 posts identified as non-/war related for the time before the war and within war. Furthermore, we focus on non-/deletions---resulting in 750 labels of which 94.7\% were consistent between non-/expert coders.

\noindent\textbf{Accuracy Evaluation.}
Table \ref{tab:accuracy-metrics} shows the evaluation results of the classifier. Our keyword-matching approach overall works surprisingly well at a precision of 0.97, with an F1 score of 0.85. It can be observed that our method prioritizes precision, \ie more than 97\% messages classified as war-related are indeed war-related. Also, we observe increased figures of false negatives (\ie lower recall), as expected.

\begin{table}[]
\centering
\caption{Accuracy metrics for the resulting war-related content classifier.}
\begin{tabular}{lrrrr}
\toprule
\textbf{Dataset} & \textbf{Accuracy} & \textbf{Precision} & \textbf{Recall} & \textbf{F1} \\
\midrule
Tripadvisor      & 0.8523            & 0.9565             & 0.6735          & 0.7904      \\
Google Maps      & 0.9231            & 0.9910             & 0.8594          & 0.9205      \\
VK               & 0.8462            & 0.9762             & 0.7193          & 0.8283      \\
\midrule
Total            & 0.8752            & 0.9773             & 0.7588          & 0.8543      \\
\bottomrule
\end{tabular}%
\label{tab:accuracy-metrics}
\end{table}

\begin{figure}[t!]
  \centering
  \begin{subfigure}[b]{\columnwidth}
      \centering
      \includegraphics[width=\columnwidth]{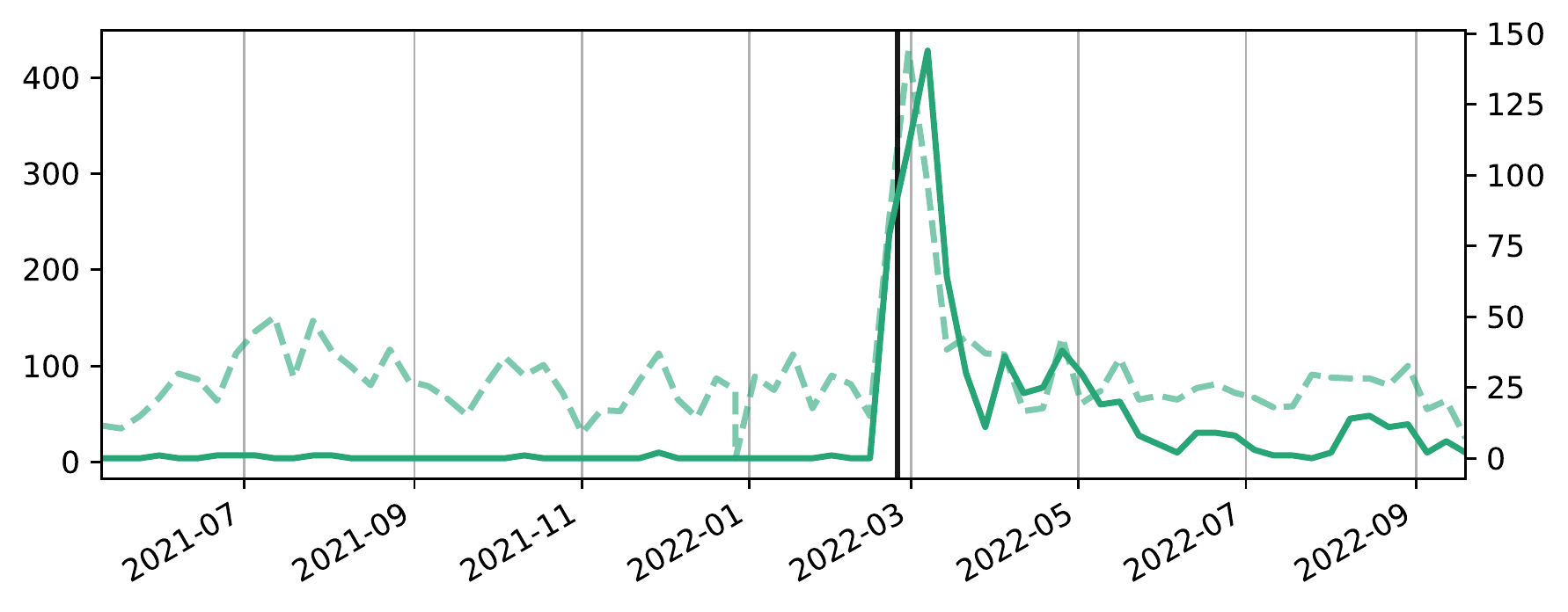}
      \vspace*{-2em}
      \caption{Tripadvisor}
      \label{fig:volumes:ta}
      \vspace*{0.5em}
  \end{subfigure}
  \begin{subfigure}[b]{\columnwidth}
      \centering
      \includegraphics[width=\columnwidth]{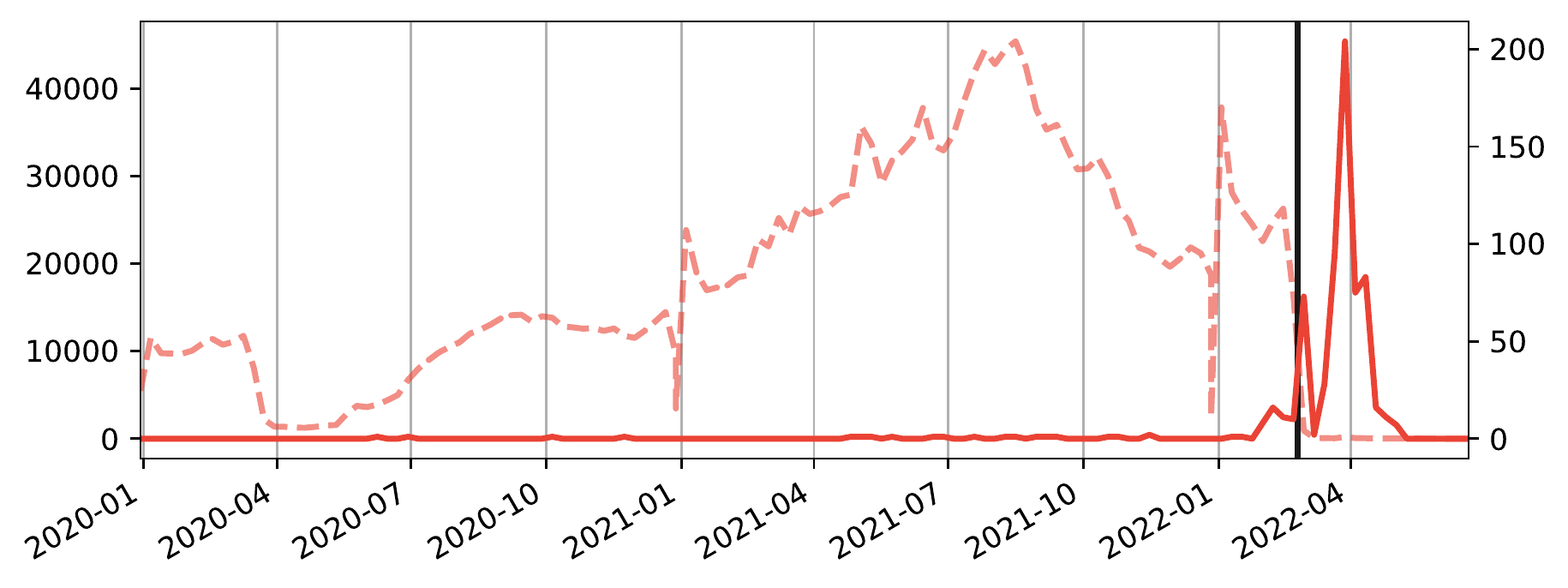}
      \vspace*{-2em}
      \caption{Google Maps}
      \label{fig:volumes:gmaps}
      \vspace*{0.5em}
  \end{subfigure}
  \vspace*{-2.2em}
  \caption{Volume of weekly published posts and reviews per dataset. Dashed lines show non-war messages (left axis).}
  \label{fig:volumes}
\end{figure}

\noindent\textbf{Activity on Tripadvisor and Google Maps.}
Figure~\ref{fig:volumes} shows the weekly rate of war and non-war related messages in both Tripadvisor and Google Maps. In the previous section we noted an increased amount of activity Tripadvisor. Now, we observe that this increase was done mostly due to war-related content. We note that, due to the conversational nature of Tripadvisor's forum threads, some messages, even if not classified as war related, were replies to previous war-related messages. Also, during the first two months right after the conflict began, we see a similar pattern for both war and non-war related, whereas from early May, the number of weekly war-related content decreased, with weeks with less than 10 messages related to war.
In the case of Google Maps, our volume analysis showed a steep drop in the number of messages, which is attributed to the active removal (first) and complete blocking (afterwards) by Google of reviews in places from Russia. An interesting pattern is that, even with such a drop, we observe that most of the messages posted after Feb 24, 2022 in Google Maps were war-related, with a peak of nearly 200 messages in the first week of March. This coincides with the beginning of our data collection.

Our results confirm that Tripadvisor and Google Maps were used as effective side channels to avoid state information controls and to communicate war-related content, possibly because of the blocking of other platforms.

\begin{figure}[t!]
    \centering
    \includegraphics[width=\columnwidth]{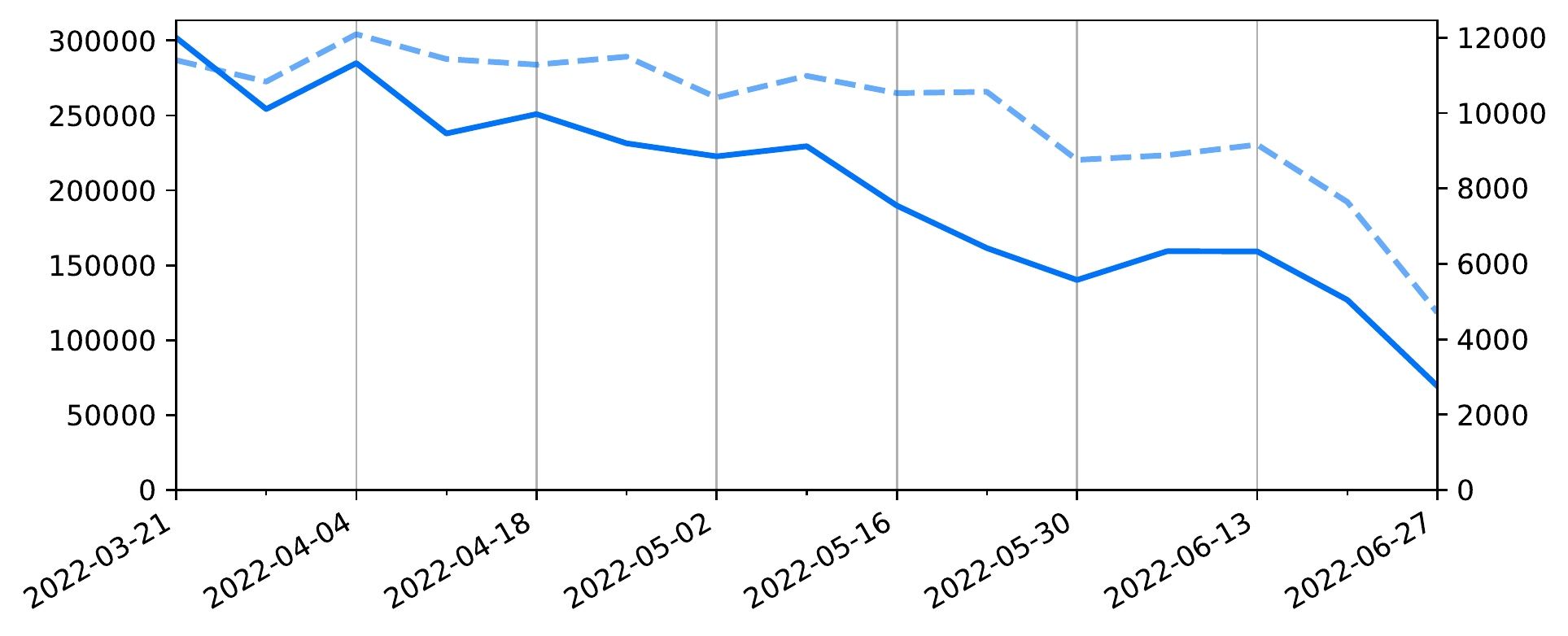}
    \vspace*{-2em}
    \caption{
        Volume of weekly published posts and reviews in VK. Dashed line shows non-war messages (left axis).
    }
    \label{fig:vk-volume}
\end{figure}

\noindent\textbf{Activity on VK.}
We analyzed activity (especially discussions in relation to the war) in VK as a complement to the insights gained when looking at the use of Tripadvisor and Google maps as side communication channels.
Figure~\ref{fig:vk-volume} presents the amount of (conservatively) identified war-related posts within the VK dataset over time. Note that we only show the period that has been actively crawled.
The number of observed posts decreases over time, of which war-related posts account for $\approx$3.1\% according to our conservative keyword approach---naturally indicating ongoing public discourse.
From a qualitative perspective, most contents are largely in line with Russian propaganda; we also find evidence of more critical opinions---or likewise posts trying to inform the Russian population about the war.
Noteworthy, we find banned words by the Russian government, under threat of fine, in 2,813 instances (1,597 \textit{``assault''}, 1,117 \textit{``invasion''}, 99 \textit{``declaration of war''}), whereas \emph{``special military operation''} appears just 179 times.

\noindent\textbf{VK Content Popularity.} 
As any Online Social Network (OSN), VK offers user-engagement metrics to infer content popularity like views, likes and shares.
For views, we observe a heavy-tailed distribution common within OSN.
Comparing war with non-war related threads (initial posts), we observe equal amounts in views on average, while likes and shares of war-related posts appear more frequent in the lower quantiles.
The same holds true for replies, while the war-related content appears less skewed in the upper percentiles.
Focusing on thread lengths, we observe a huge shift towards longer discussions for war-related content.
We conclude that content receives equal visibility, whereas war-related content is discussed by users more extensively.

\subsection{Topic Analysis}
\label{subsec:topic-analysis}
We perform an analysis of the topics mentioned in Tripadvisor posts and Google Maps reviews to understand the difference in discourse before and after the war. The VK dataset is excluded from this comparative analysis as we lack historical data for this platform.
We determine the topics of a message based on the keywords it contains. We manually group all the war-related keywords (listed in Appendix~\ref{appendix:keywords} for reference) into different classes to come up with a list of 10 topics. This process was conducted by two expert coders.

\begin{figure}[t!]
  \centering
  \begin{subfigure}[b]{\columnwidth}
      \centering
      \includegraphics[width=\columnwidth]{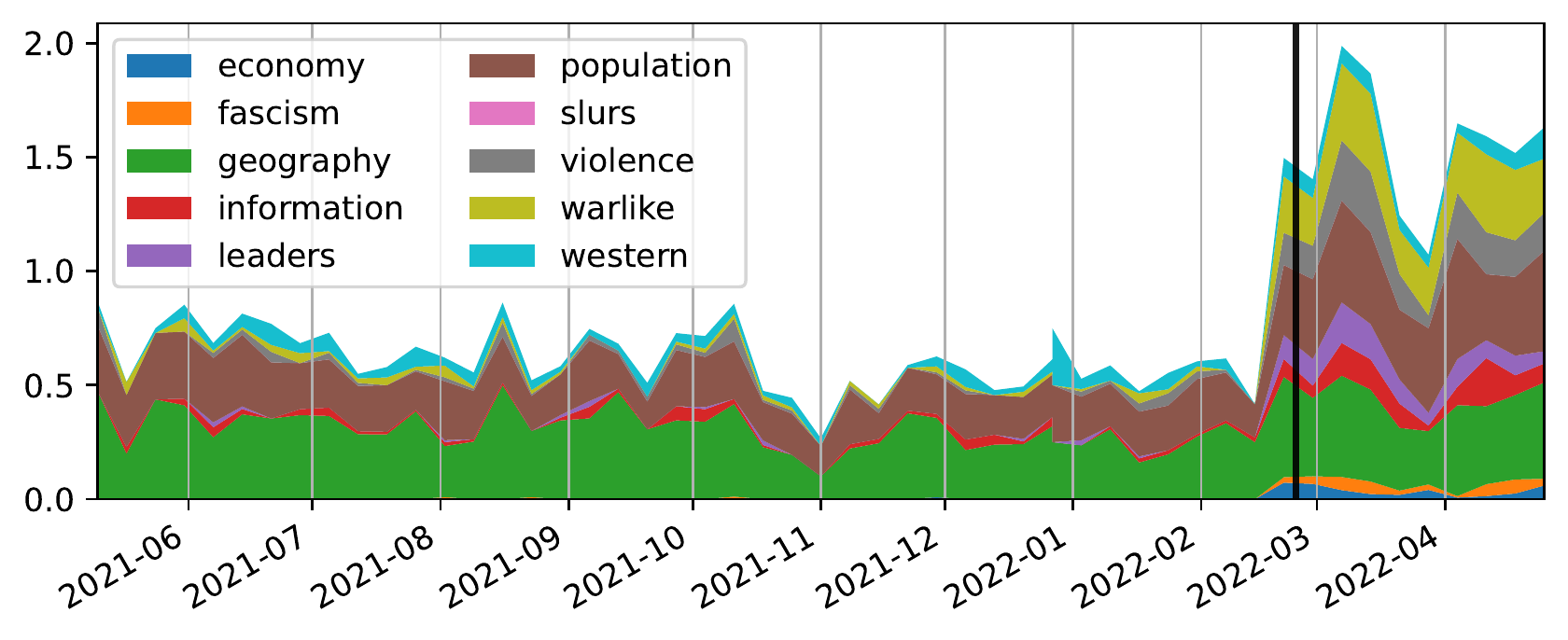}
      \vspace*{-2em}
      \caption{Tripadvisor}
      \label{fig:topics:ta}
      \vspace*{0.5em}
  \end{subfigure}
  \begin{subfigure}[b]{\columnwidth}
      \centering
      \includegraphics[width=\columnwidth]{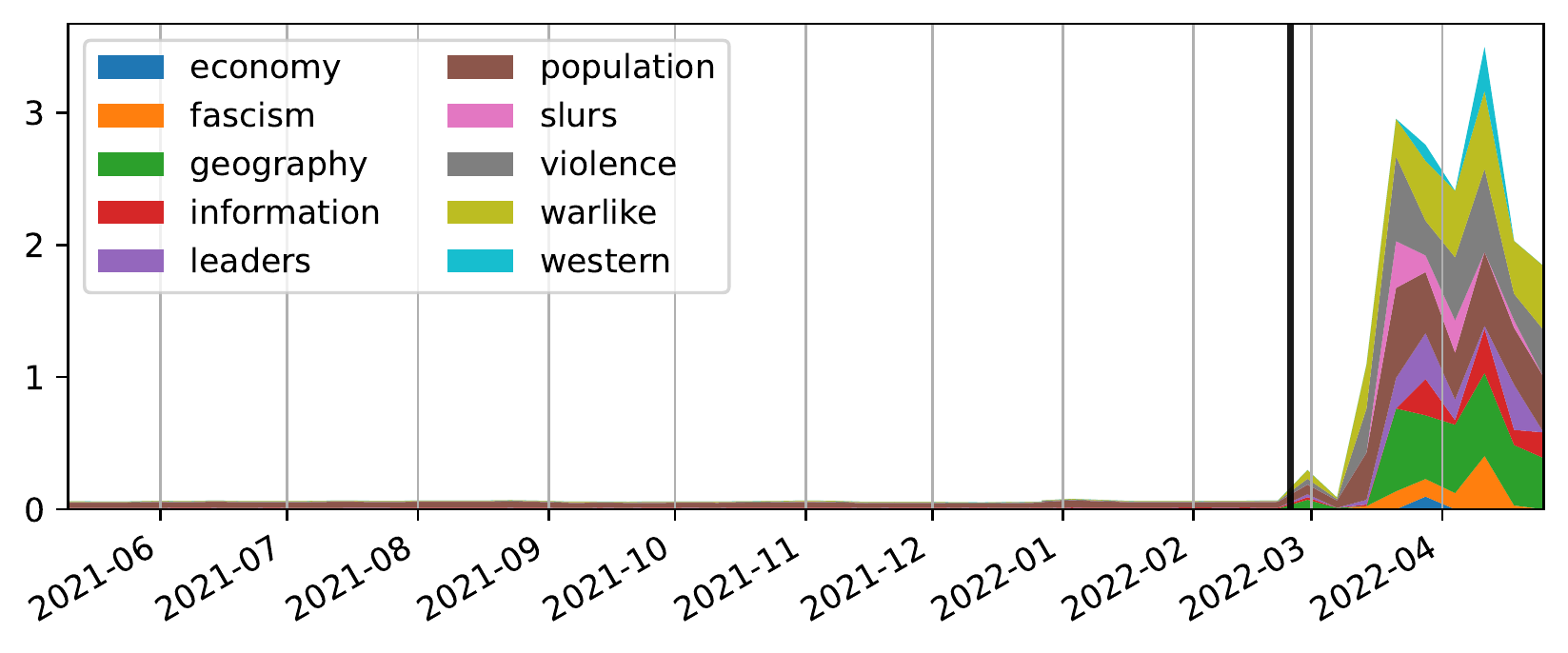}
      \vspace*{-2em}
      \caption{Google Maps}
      \label{fig:topics:gmaps}
  \end{subfigure}
  \vspace*{-2em}
  \caption{Evolution of the ratio of topics found in posts and reviews per week. The bold vertical line indicates the beginning of the Russian invasion.}
  \label{fig:topics}
\end{figure}

Figure~\ref{fig:topics} shows the change over time in the ratio of topics for the mentioned platforms. The number of messages is normalized to account for variations in volume.
In the case of Tripadvisor, we observe that topics like \textit{warlike}, \textit{violence} or \textit{fascism} gained popularity immediately after the beginning of the war. Similarly, Google Maps reviews posted before the war do not tend to mention any of the identified topics and focus almost exclusively on user recommendations. We observe a radical change in the discourse shortly after March 2022, correlating with the addition of war-related reviews to the platform.

\subsection{Targeted Places in Google Maps}
\label{subsec:geography}
\begin{figure}[t!]
    \centering
    \includegraphics[width=\columnwidth]{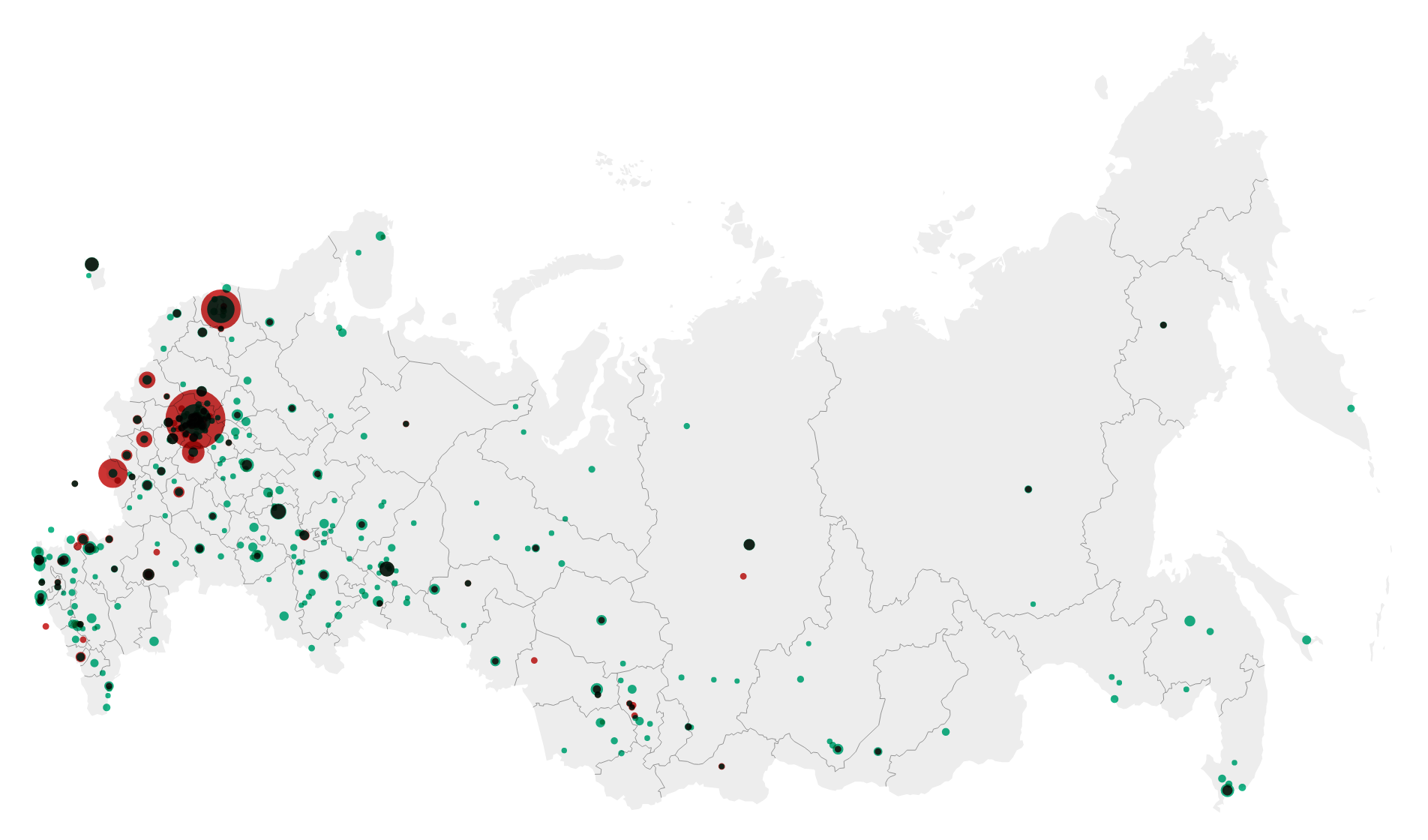}
    \caption{Ratio of war-related and non war-related reviews per Russian town (in \textcolor{purple}{red}), non-war-related reviews (in \textcolor[rgb]{0.01,0.68,0.49}{green}) and both (in \textcolor[rgb]{0.1,0.1,0.1}{black}).}
    \label{fig:gmaps-places}
\end{figure}

We studied which cities and regions tend to be the objective of war-related reviews. Unfortunately, this analysis is only possible on Google Maps, as it is the only data source associated with locations in Russia. We observe that 108 towns over a total of~\gmapsTowns found on Google Maps contain at least one war-related review. These towns cover significant areas of Russia, including cities like Krasnoyarsk in Siberia, Vladikavkaz in the Caucasus, and Moscow in the central district (Figure~\ref{fig:gmaps-places}). We observe that half of the total war-related reviews concentrates on Moscow and St. Petersburg, with 37\% and 15\% respectively. However, the ratio between war and non-war reviews on cities close to the Ukrainian border or around Moscow is higher than in other regions (red dots in Figure~\ref{fig:gmaps-places}). Some examples are Belgorod (frontier city) with 0.6\% of all the reviews being war-related ones or Ryazan (southeast of Moscow) with 0.25\%, while in cities such as St. Petersburg drops to 0.06\%.

\takeaway{
We observe substantial changes in the number of daily posts and reviews since the beginning of the war.
These correlate with the blocking of major social platforms and news sites in Russia, and with a call by activist groups to reach out to the Russian population.
Our automatic content labeling reveals that the traffic increase is mostly due to war-related content, and topic analysis confirms a noticeable change in user discourse.
These findings ratify that Tripadvisor and Google Maps were used as side channels to disseminate war-related information to Russian citizens, possibly because of state-level blocking of other platforms.
}

\vspace{-1mm}
\section{Intent and Purpose of Content}
\label{sec:case-studies}

Section~\ref{sec:contentanalysis} offered a high-level overview of the main war-related topics being discussed using qualitative analysis. We next describe the 4 main topics found: $(i)$ disinformation targeting Russian citizens, $(ii)$ humanitarian aid, $(iii)$ polarization, and $(iv)$ travel advice. We provide a general discussion on the topics, showing some of the sample messages verbatim.

\vspace{1mm}
\subsection{(Dis)information and Censorship Bypass}
Our analysis suggests that the three analyzed platforms have been used as a channel to fight against Russia's information war because they were not blocked by Russian authorities---though, as shown in Section~\ref{sec:moderation}, their content was moderated by administrators. Thus, most of the reviews found in Google Maps and the posts from Tripadvisor seek to inform Russian citizens about the war (see examples from Appendix~\ref{appendix:samples}).

These messages are often orchestrated as organized campaigns, as discussed in \ref{sec:org_campaigns}. For example, the following message appeared in 52 different reviews in Google Maps, written both in English and Russian:
\begin{quote}
    BE AWARE!!!! MURDERERS! THEY (RUSSIANS) ARE KILLING KIDS, SENIORS AND WOMEN IN MARIUPOL AND MANY OTHER CITIES IN UKRAINE!!!!! SHAME!!! DON'T BUY THEIR ROTTEN FOOD!!!
\end{quote}

As part of this information war, one Tripadvisor user claimed on the earliest days that \textit{``[Tripadvisor] has been the target of a team of Kremlin trolls who are paid to try to control the narrative on here around the current state of Russia.''}
In this post, the user asked the community for \textit{``comments correcting falsehoods''} and pledge for \textit{``cooperation between internet users and observers who are able to expose and compromise trolls,''}, also asking for platform moderation due to the aggressive speech \textit{``extensively employed by trolls.''}
This discourse is often repeated in the platforms, and the intent is to inform Russian citizens about the actual events happening:
\begin{quote}
    Much Fake news is spreading in the official Russian Press Media. [..]. Be careful with the Russian agents that will be paid by the Putin government, some of them are trolls of the KGB. They are rats sewer who help Putin spread false information, be careful.
\end{quote}

We also observe messages assisting Russians to bypass censorship. For example, a message with instructions on \textit{``how to get around restrictions on BBC services in Russia''} was posted by the same user in 8 threads simultaneously.

We also identified several pro-Russian users who also engaged in these efforts. For instance, one user in Tripadvisor claiming that \textit{``any message that does not fit into your picture of the world is automatically branded with the phrases `your garbage', `nonsense', `propaganda'''}, or a review in Google Maps, which was repeated in 22 different places, that claims that \textit{``Moscow already knows that the dill [a Russian language ethnic slur to refer to Ukrainians] will hit''} and \textit{``we are left here like a live shield and do not give information, they want bombs for the picture to fall on us''}, finally claiming that this is \textit{``information from a very reliable person in the army''}.

Overall, propaganda and disinformation messages were one of the main topics of discussion observed.

\subsection{Humanitarian Help}
Another frequent topic of discussion is humanitarian help and advice. Due to these platforms being accessible worldwide, people decided to use them to inform and raise awareness about the humanitarian needs for Ukrainian refugees. Users both asked for humanitarian help, and also provided advice to Ukrainian refugees in various aspects. These include cheap or free options to get out of the country (\textit{``Free travel on [redacted] flights for Ukrainians''}), or free accommodation in hotels and apartments abroad (\textit{``An independent platform connecting Ukrainian refugees with hosts who can offer free housing''}). Also, various users promote Non-Governmental Organizations that are helping refugees, such as Razom or Redcross (\textit{``Please ask your rugby club to donate to Razom for Ukraine.''}), and informing on charities helping kids (\textit{``A friend is at present involved in getting orphaned kids out of Dnipro [...] The charity, Edinburgh based Dnipro Kids' is hoping to get many more children to safety''}).

\vspace{1mm}
\noindent\textbf{Polarization.}
Besides the main topics discussed, we observe one common pattern common to the three studied platforms: there is a high level of polarization. There is a clear positioning pro-war and anti-war (predominantly the second). This leads to various instances of hate speech that we will not reproduce in this study. As discussed in Section~\ref{sec:moderation}, TA actively banned such content and removed (at least) 191 posts, including entire threads, when these turned into somehow aggressive discussions between members. 
Also, as we discussed in Section~\ref{subsec:topic-analysis}, we observe a large increase in messages that include violent and offensive words (\eg killer, troll, monster, or nazi) since the beginning of war. Therefore, we observe how platforms that are initially intended for licit purposes (\eg provide travel advice or offer customer reviews), have been radicalized to an unprecedented extent.

\vspace{1mm}
\noindent\textbf{Travel advice.}
The final common topic which we have observed, mostly in Tripadvisor, is about assistance for travelling in and out of Russia and Ukraine. Even if this is expected---this was the original purpose of enabling war-related content in the forums---we observe that sometimes the conversations are easily polarized towards one side. For example, one user informed about the potential danger in traveling between Moldova and Odessa (\textit{``I recommend refraining from traveling in this direction, Zelensky's soldiers will be happy to use you for another provocation.''}). However, these are isolated cases, and most of the advice is done for refugees willing to leave the country (\textit{``Tomorrow will be a bus (for free) in Lviv. Bus will be waiting for children and women next to main train station in Lwiw''}).

\subsection{Orchestrated Content Dissemination}\label{sec:org_campaigns}

We identified instances of the same slogans being posted by different users on the same platform and even across platforms. 
Given the public calls made by activist groups to inform Russian citizens about ongoing war events, we analyzed if the posted messages contain organized campaigns (\ie spamming the same text) in addition to personal or individual statements.
We identify several identical posts on Tripadvisor and Google Maps with patriotic statements
or informing Russian citizens about the war pledging for an end.
For Google Maps, we identify 8 posts each being replayed more than 10 times, totalling 188 instances. The same holds for Tripadvisor, though the amount in this case is smaller: 7 posts posted 17 times. In the case of VK,
which is dominated by Russian users, we find 28 war-related messages posted 399 times.
As discussed before, our qualitative look into the contents provides a wide spectrum of pro-Russian and pro-Ukrainian messages, whereas others primarily pledge to stop the war.

\begin{figure}[t!]
  \centering
  \begin{subfigure}[b]{\columnwidth}
      \centering
      \includegraphics[width=\columnwidth]{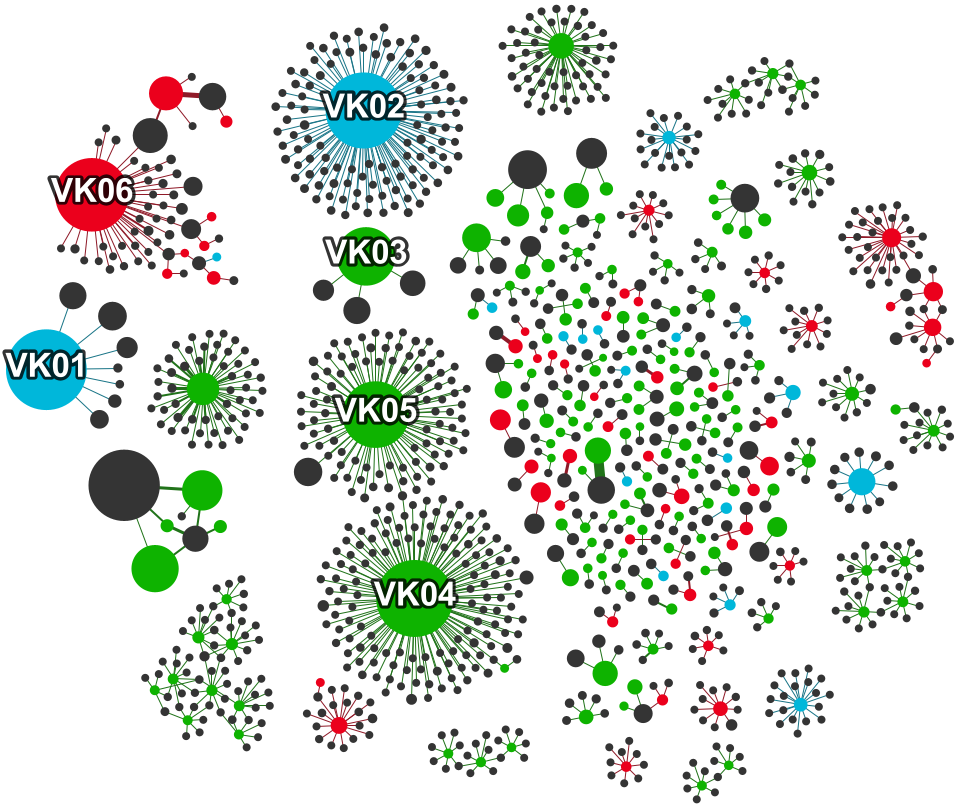}
      \vspace*{-1.1em}
      \caption{VK}
      \label{fig:campaigns:vk}
      \vspace*{0.5em}
  \end{subfigure}
  \begin{subfigure}[b]{\columnwidth}
      \centering
      \includegraphics[width=0.5\columnwidth]{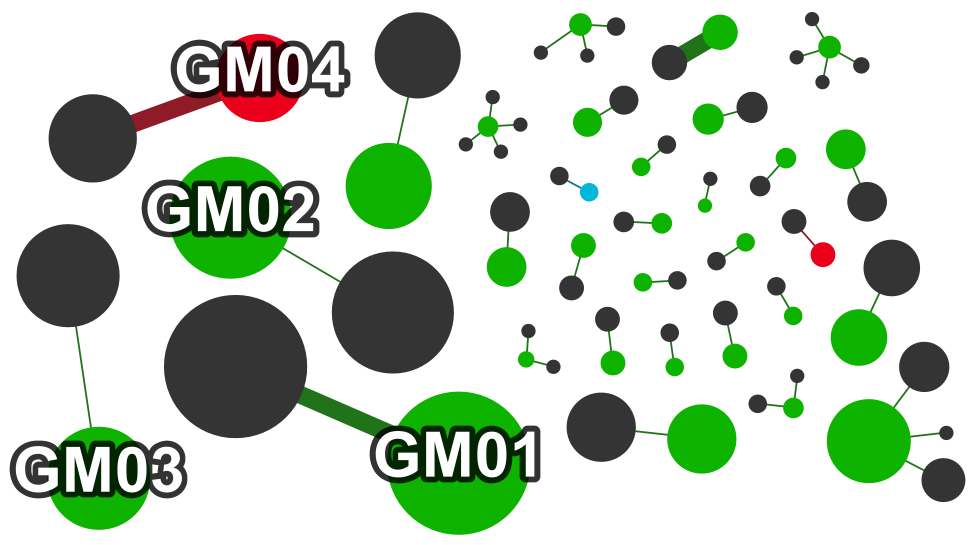}
      \vspace*{-0.4em}
      \caption{Google Maps}
      \label{fig:campaigns:gmaps}
  \end{subfigure}
  \vspace*{-2em}
  \caption{Network graphs of the largest campaigns observed. Nodes in \textcolor[HTML]{eb001c}{red} show pro-war campaigns; in \textcolor[HTML]{0eb300}{green} against-war campaigns; and in \textcolor[HTML]{00b7da}{blue} neutral or inconclusive. \textcolor[HTML]{343434}{Gray} nodes represent users.}
  \label{fig:campaigns}
\end{figure}

Figure~\ref{fig:campaigns} shows the top largest campaigns found in VK and Google Maps by volume of messages, alongside the users that took part in them. We provide examples in Appendix~\ref{appendix:samples} for the campaign nodes that have a label.
Duplicated content in VK is mostly composed of messages leaning towards an anti-war stance. These campaigns are fueled by individual accounts that post a single message and do not participate in other known campaigns. To a lesser extent, we also find smaller isolated networks of duplicated content in which users participate by posting messages belonging to two or more different campaigns.
In the case of Google Maps, campaigns are almost exclusively posted by a single user that distributes their messages across different places.

Finally, we also found patriotic and war-related slogans distributed across Google Maps and VK. Examples are ``\foreignlanguage{russian}{Слава Украине}'' or ``\foreignlanguage{russian}{Слава Украины}'' (meaning ``Glory to Ukraine'' in Russian); ``\foreignlanguage{russian}{Слава Україні}'' (meaning ``Glory to Ukraine'' in Ukrainian); and ``\foreignlanguage{russian}{Нет войне}'' (meaning ``No War'' in Russian) or ``Stop War''. These slogans are commonly used as a closing remark within a larger message.

\takeaway{
Tripadvisor and Google Maps were used to disseminate messages about: $(i)$
(dis)information targeting Russian citizens, $(ii)$ humanitarian aid, $(iii)$ hate speech, and $(iv)$ travel advice. We found evidenced of information campaigns supported by multiple users who contributed to disseminating both pro-Ukraine and pro-Russia messages and slogans.
}

\section{Platform Moderation}\label{sec:moderation}

Platform operators reacted differently to the use of their services as side channels for disseminating war-related information. We next discuss content moderation---or lack thereof---in the three platforms analyzed in this study. For reference, we first analyze removal of posts by administrators, as observed during our daily crawls. We restrict this analysis to the period when we conducted hourly crawls, \ie during the first two months of the conflict. We note that measuring VK post removals is not feasible due to crawler limitations (\S\ref{subsec:crawl-limitations}).
In the case of Google Maps, we estimate removals based on the latest time that our crawler recorded a given review. This yields good results as we daily crawl all reviews (new and previously existing ones) for all monitored places.
In the case of Tripadvisor, a removed post is replaced by a placeholder message from the admins indicating the reason for removal. However, the timestamp of this replacement is not provided. For Google Maps, we estimate the removal date as the first timestamp where we observe that a message was replaced by the placeholder message. We conduct hourly crawls for the analyzed period so that we can measure platform moderation with one-hour precision.

\begin{figure}[t!]
    \centering
    \includegraphics[width=1\columnwidth]{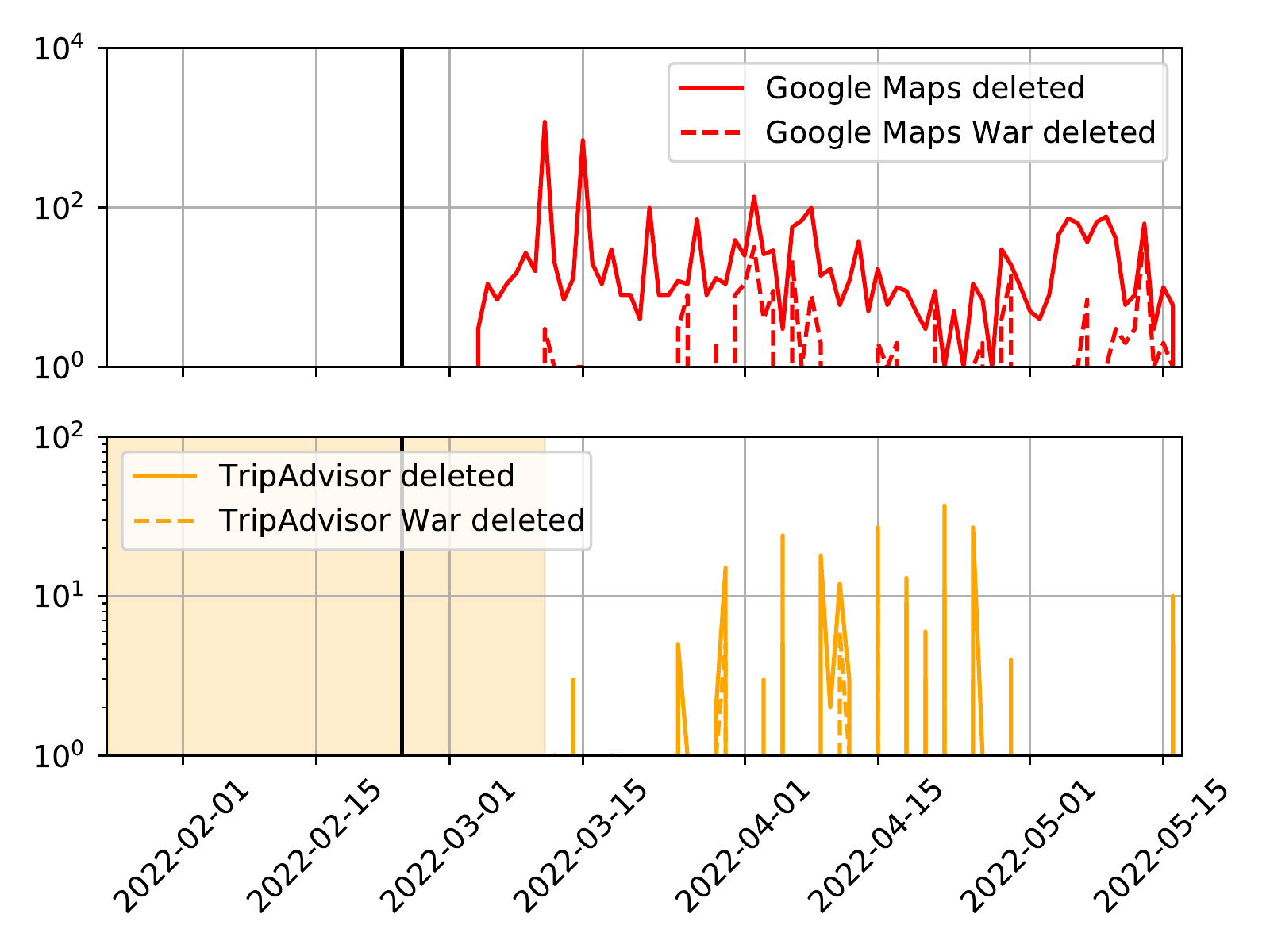}
    \vspace*{-2em}
    \caption{
        Number of daily posts removed from Google Maps (top) and Tripadvisor (bottom), and the number of them labeled as ``war related''.
    }
    \label{fig:removals-labeled}
\end{figure}

Figure~\ref{fig:removals-labeled} shows the number of war and non-war deleted posts for both Google Maps and Tripadvisor. We observe that the volume and also the frequency of removals is higher for Google Maps than Tripadvisor, as opposed to the new entries where we observe similar patterns.
This suggests that both sites implemented different content moderation policies during the war-time. We next discuss each platform moderation and reasons for removals, when available.

\vspace{1mm}
\noindent\textbf{Tripadvisor. \xspace}
There is evidence of various posts (and entire threads) being removed by forum administrators. Table~\ref{tab:removalTripadvisor} shows the number of posts and the reason provided in the placeholder message left. We narrow down our analysis on messages removed after the war started. Note that up to 121 posts were removed at the author's request according to the metadata offered by the platform. Hate speech and harassment are the two more prevalent categories when posts are removed by the platform.
For the 34 threads (215 posts) that have been removed completely, we ignore the reasons for such removals. Out of these, 13 ($\sim$38\%) have a unique post, and 4 ($\sim$12\%) only contain one reply. Meanwhile, 7 threads ($\sim$20\%) have more than 13 replies. We confirm through manual inspection that the reasons for thread removal typically fall into two categories: (i) the conversation in the threads completely deviate from its original purpose (\eg provide objective information about the conflict) towards political discussions or even hate speech; or (ii) the thread is initiated with the sole purpose of propaganda or another advertisement, and it is removed quickly, sometimes even before it gets any reply. 
Table~\ref{tab:lifespanTA} shows the lifespan of the 69 items (\ie a post or the entire thread) removed by administrators in the period when we conducted hourly crawls (Mar 12, 2022 - May 12, 2022). Posts in general are removed faster than entire threads, but we also observe that those threads without replies (\ie containing only the OP message) are removed as quickly as regular posts. This confirms that active platform moderation is in place and that much of the content removal is linked to the war.

\begin{table}[]
    \centering
    \caption{Lifespan of content removed in Tripadvisor.}
    \begin{tabular}{lrrr}
        \toprule
               & \textbf{Threads} & \textbf{Posts} & \textbf{All} \\
        \midrule
        Total  & 31               & 38             & 69 \\
        Mean   & 7d 23h 15m       & 2d 16h 16m     & 5d 1h 19m \\
        Median & 2d 16h 44m       & 1d 14h 32m     & 2d 8h 28m \\
        Stdev  & 12d 8h 42m 40s   & 2d 14h 0m      & 8d 20h 5m \\
        \bottomrule
    \end{tabular}%
    \label{tab:lifespanTA}
\end{table}

\begin{table}
    \centering
    \caption{Reasons for content removals in Tripadvisor.}
    \label{tab:removalTripadvisor}
    \begin{tabular}{lr}
        \toprule
        \textbf{Reason} & \textbf{Posts}  \\
        \midrule
        Entire thread was removed & 215\\
        Off-topic chat & 147 \\
        Removed by author & 121\\
        Harassment to other users & 102\\
        Hate speech or inappropriate language & 89\\
        Self-promotional advertising  & 31\\
        Not written in English  & 14\\
        Copyright infringement  & 6\\
        Multi-account detected  & 2\\
        \midrule
        Total & 727\\
        \bottomrule
    \end{tabular}%
\end{table}

\vspace{1mm}
\noindent\textbf{Google Maps.\xspace}
We find evidence of platform moderation for Google Maps, as war-related
reviews have a much shorter lifespan than the rest, typically lasting less than 50 days as opposed to those until the end of our crawling.
Shortly after the beginning of the war, Google Maps temporarily suspended posting new reviews on Russian places in an attempt to prevent the generation of content that violates company policies~\cite{wsjTripadvisorGoogleMaps}.
This led to the drastic reduction on the number of daily published reviews, as shown in Figure~\ref{fig:volume-all}.
While war-related content is not explicitly prohibited in Google Maps, Google alleged that these reviews were considered ``off-topic,'' a category that is prohibited in Google Maps, justifying their temporal suspension~\cite{googleMapsContentPolicy}.
According to our data, this temporary banning was still active by May 2022---just 8 posted reviews per day. Nevertheless, we find 18 war-related reviews that somehow bypassed Google Maps' moderation.

\takeaway{
There is evidence of platform operators actively removing messages that contain war-related content. The most common reasons for content removal include off-topic message, harassment and hate speech, or more generally ``content that violates ToS.'' The lifespan of moderated content ranges from a few hours to several weeks, with some messages escaping moderation.
}

\section{Conclusions}

This paper explores for the first time how Internet services such as Tripadvisor and Google Maps are used to bypass state-level censorship.
Specifically, we empirically study how these services were used to deliver information related to the Russian invasion of Ukraine. Using a dataset collected during the first weeks of the war, we observe the shifts in the pattern of daily user post volume and removals, and also duplicated content suggesting intentional and organized campaigns to disseminate such information. Our content analysis using both quantitative and qualitative methods confirms that there is indeed a peak in war-related narratives in the reviews in these platforms. For comparison, we observe a similar pattern in the Russian social network VK.

The use of place and business reviews as side channels forced platforms to apply content moderation policies. In the case of Tripadvisor, our analysis suggests that administrators do perform intensive content moderation. However, they allow (and indeed encourage users) to discuss and inform about the conflict, mostly to provide information about safe traveling in and out Russia and Ukraine. In the case of Google Maps, there is also evidence of bulk removals leading to a temporary suspension of the reviewing activity that extends up to the time of this writing.

Overall, our study reveals how three unblocked online platforms were explored to circumvent state-wide information controls. However, its contributions go beyond the analysis of an interesting phenomena on the Internet, as it provides new insights on human behavior displacement in times of crisis, and raises the question on the role of the Internet in these periods and the effectiveness of Internet censorship.

\begin{acks}
This project has been funded by
the Spanish grants ODIO (PID2019-111429RB-C21 and PID2019-111429RB-C22), the Region of Madrid grant CYNAMON-CM (P2018/TCS-4566), co-financed by European Structural Funds ESF and FEDER, and by European Research Council (ERC) under Starting Grant ResolutioNet (ERC-StG-679158).
The opinions, findings, and conclusions or recommendations expressed are
those of the authors and do not necessarily reflect those of any of the
funding agencies.
\end{acks}

\bibliographystyle{ACM-Reference-Format}

\balance
\appendix

\section{Examples of posts}
\label{appendix:samples}

Table~\ref{tab:examples} contains illustrative examples of war-related posts and reviews we found across the three studied platforms.

\begin{table*}[!bpt]
\caption{Example posts and reviews. Translated texts appear in \textit{italic}.}
\label{tab:examples}
\centering
\begin{tabularx}{\textwidth}{rX}
\toprule
\textbf{ID} & \textbf{Content} \\
\midrule

\textcolor{taColor}{TA01} & PUTIN is killing is killing innocent Ukrainians; men women, children and their pets. Please help. \\
\midrule
\textcolor{taColor}{TA02} & Here's a reminder of how to get around restrictions on BBC services in Russia: Download the Psiphon app from the AppStore or Google Play Store - Look for the dedicated BBC site on the Tor Browser which can be found using this URL. \textcolor{gray}{[...]} \\
\midrule
\textcolor{taColor}{TA03} & \textcolor{gray}{[NICKNAME]}, I don't like to speak about political opinions, Kills thousand of people in Ukrania and thousand of Russian young soldiers are not acceptable in this century, I can not speak about tourism and forget the atrocities in this crazy war. \textcolor{gray}{[...]} \\
\midrule
\textcolor{taColor}{TA04} & Since you want to turn this thread into politics you might want to check and find out that Ukraine killed 14000+ innocent civilians -Ethnic Russians in Donbass, Eastern Ukraine in 2014-2022 and it keeps shelling this region to cause them to die or leave. Why everyone thinks this was/is OK? And noone talks about it. \\
\midrule

\textcolor{gmapsColor}{GM01} & \textit{Your children, relatives are being mobilized. Do you want to find them on our lists? We don't. How many of your soldiers died? - THOUSAND?! Don't believe it! - Already over 17k dead Russian soldiers in Ukraine!!! - All Actual Information about the Fallen Russian Armed Forces in Ukraine is here: - https://t.me/rf200\_now - https://youtube.com/c/VolodymyrZolkin - We are a humanitarian project to inform the relatives of those killed about their fate WATCH, LISTEN, THINK, ANALYZE - IT IS IMPOSSIBLE. YOUR AUTHORITIES, THE MEDIA LIE TO YOU...} \textcolor{gray}{EMOJI} \\
\midrule
\textcolor{gmapsColor}{GM02} & BE AWARE!!!! MURDERERS! THEY (RUSSIANS) ARE KILLING KIDS, SENIORS AND WOMEN IN MARIUPOL AND MANY OTHER CITIES IN UKRAINE!!!!! SHAME!!! DON'T BUY THEIR ROTTEN FOOD!!! \foreignlanguage{russian}{ВНИМАНИЕ!!! УБИЙЦЫ! ОНИ (РАШИСТЫ) УБИВАЮТ ДЕТЕЙ, СТАРИКОВ И ЖЕНЩИН В МАРИУПОЛЕ И МНОГИХ ДРУГИХ ГОРОДАХ УКРАИНЫ!!!!! НЕ ПОКУПАЙТЕ ИХ ГНИЛУЮ ЕДУ, ИХ РУКИ В КРОВИ!!!} \\
\midrule
\textcolor{gmapsColor}{GM03} & I am an American and I am writing about Ukraine. If the only news you receive in Russia is from your state media, you are being told lies. It is critical that you know the truth about the Russian invasion of Ukraine. The Russian pretext of restoring peace to Ukraine and removing Nazis from the government is a lie. \textcolor{gray}{[...]} \\
\midrule
\textcolor{gmapsColor}{GM04} & \textit{Ukrops have already been given many offensive weapons concealed!!!! A strike on Belgorod is being prepared. So far, there is no exact date when it will happen. Moscow already knows that Ukrop will strike, they are already digging trenches in Shebekino!!!!} \textcolor{gray}{[...]} \\
\midrule
\textcolor{gmapsColor}{GM05} & \textit{The food was great! Unfortunately, Putin has ruined our appetites by invading Ukraine. Stand up to your dictator, stop killing innocent people! Your government is lying to you. Stand up!} \\
\midrule

\textcolor{vkColor}{VK01} & \textit{Relatives of soldiers killed in Ukraine demand compensation of 10 million rubles from Putin} \\
\midrule
\textcolor{vkColor}{VK02} & \textit{The 26th day of the war. There is no progress in the negotiations. While the leaders of the countries are trying to find a solution that would satisfy both sides, people continue to die. Komsomolskaya Pravda wrote, citing the AFU General Staff, that from the beginning of a special military operation in Ukraine to March 20, the Russian Armed Forces lost 96 airplanes, 118 helicopters and 14,700 servicemen.} \textcolor{gray}{[...]} \\
\midrule
\textcolor{vkColor}{VK03} & \textit{Ivan Ivanov from Chelyabinsk recorded a video appeal to Putin V.V. with a demand to pay compensation of at least 10 million rubles for his uncle - serviceman Ilyas Rifovich Fahretdinov, born on June 17, 1992 (Warrant Officer) from Novogorny settlement, Chelyabinsk region, who died during the special operation in Ukraine. (Don't be frightened, they are really relatives). A nephew of the person who died is against the special military operation, against Putin's regime in Russia.} \\
\midrule
\textcolor{vkColor}{VK04} & \foreignlanguage{russian}{Слава Украине!!!} \textcolor{gray}{EMOJI} \\
\midrule
\textcolor{vkColor}{VK05} & \foreignlanguage{russian}{Слава УКРАЇНІ} \textcolor{gray}{EMOJI} \\
\midrule
\textcolor{vkColor}{VK06} & \textit{Vladimir: Since some of the assets of the Central Bank of Russia were arrested, Russia has the right not to comply with the American financial model anymore. I suppose Putin deliberately allowed some of our Central Bank's assets to be seized in order to get an excuse not to comply with the model created by the U.S. Federal Reserve.} \textcolor{gray}{[...]} \\
\midrule
\textcolor{vkColor}{VK07} & \textit{https://blagotvorec.org/ - A volunteer organization comprised of people who want to help all those affected by the war in Ukraine in a difficult life situation} \\

\bottomrule
\end{tabularx}
\end{table*}

\section{War-related Keywords}
\label{appendix:keywords}

\begin{table*}[!bpt]
  \centering
  \caption{List of keywords used for classifying posts and reviews.}
  \label{tab:classifierKeywords}
  \small
  \begin{tabularx}{\textwidth}{rX}
      \toprule
      \textbf{Weight} & \textbf{Keywords} \\
      \midrule
      3 & airstrikes, armed, army, attack, attacked, attacks, bomb, bombard, bombarded, bombarding, bombed, bombing, bombs, bullets, conflict, dead, death, deaths, dictator, die, died, dying, invade, invaded, invades, invading, invasion, kill, killed, killer, killers, killing, massmurder, military, missile, missiles, murder, murdered, murderer, murderers, murdering, murders, rashists, rospropaganda, russofascist, russofascists, shoot, shooting, shot, soldier, soldiers, tanks, troops, war, weapon, weapons, wwiii, \foreignlanguage{russian}{армией, армии, армия, армія, бандера, бандеровское, бандеровцев, бомбили, бомбить, бомбы, бригада, взрывов, взрывчаткой, взрывы, военная, военное, военной, военную, военные, военным, военных, война, войне, войной, войну, войны, войск, войска, всу, вторгается, вторгаться, вторгаются, вторглась, вторгся, вторгшись, вторжение, выстрелов, гибнут, дітей, днр, застрелен, напали, обеспеченная, оружие, оружием, оружия, погибли, пулевые, путлер, путлеру, разбомбив, разбомбили, ракет, ракетами, ракеты, рашисты, роспропаганды, русофашист, русофашистом, руссофашисты, смертей, смертельная, смерти, смерть, солдат, солдата, солдатам, солдати, солдаты, стреляют, танках, танки, танков, трупы, убивает, убивали, убивать, убивают, убийств, убийства, убийство, убийца, убийцы, убили, убит, убитий, убитых, убой, укропам, укропы, умирать} \\
      \midrule
      2 & aggression, aggressor, atrocities, \foreignlanguage{russian}{aгрессия}, blood, bodies, censorship, \foreignlanguage{russian}{chwała}, coldblooded, crime, crimes, criminal, criminals, defend, denazification, destroy, destroyed, destroying, destruction, fascist, fascists, fight, fighting, fled, flee, fleeing, genocide, guns, hate, hell, hitler, holocaust, lie, lies, lying, monster, nazi, nazis, propaganda, putin, rape, raped, raping, rapists, refugee, refugees, sanctions, tortured, traumatized, ukraina, ukraine, ukrainian, ukrainians, ukrainie, ukranian, ukranians, ukrenians, zelensky, \foreignlanguage{russian}{беженец, беженцы, боролись, воевать, воюют, выйдут, геноцид, гитлер, гитлера, денацификацией, диктатора, драться, зеленский, зеленского, зеленський, казнь, кндр, конфликт, конфликте, крови, могилку, нацист, нацистов, нацисты, обстреливаем, огонь, оккупации, окопы, олигархов, операции, операцию, операция, освободители, преступление, пропаганда, путин, путина, путину, путиным, путіна, пытали, пыток, разрушают, разрушили, резолюции, руки, санкции, спасайтесь, схватка, тероризму, украина, украине, украинец, украиной, украинская, украинские, украинский, украинским, украинских, украинского, украинской, украину, украинца, украинцев, украинцы, украины, україна, україні, українська, українським, українців, уничтожат, уничтожен, уничтожение, уничтожены, уничтожить, фашизм, фашизма, фашистами, фашистов, фашистской, фашистська, фашисты, фюрера} \\
      \midrule
      1 & american, americans, arrested, ashamed, belarus, belgorod, bless, brothers, brutal, brutally, bucha, children, cities, citizen, citizens, civilian, civilians, commiting, committed, cowardly, crimea, \foreignlanguage{russian}{cвободная}, danger, donbas, donbass, donetsk, elderly, evacuation, forces, free, freedom, government, hospitals, humanity, innocent, innocents, justice, kharkiv, kids, kindergardens, kremlin, lavrov, mariupol, mykolaiv, nato, news, nuclear, odessa, operation, peace, peaceful, people, population, president, pretext, protests, regime, resisting, restrictions, rosja, russia, russian, russians, safety, shame, shelter, special, stalin, state, stop, strike, strikes, suffering, survive, territory, truth, \foreignlanguage{russian}{ukraińcy, ukraiński}, usa, victims, violence, violent, vladimir, western, women, \foreignlanguage{russian}{азов, америка, американская, американской, американскую, америки, безопасности, белгороду, белоруссия, бешенного, біженець, біженців, білорусь, буча, варварской, вбивають, вбивство, вбивці, вбивця, вбили, вбити, війна, війни, владимир, власть, влияния, врагов, вставай, встанут, газ, газа, город, города, городам, городах, городе, городов, государства, государственные, государство, граждан, граждане, гражданин, гражданский, гражданским, гражданских, границ, державу, детей, дети, доказательств, донбасс, донбасса, донбассе, донецке, европа, европе, европы, женщин, жертвы, живой, живы, живым, жизни, жизнь, жителей, жители, жителями, жить, жінки, заживо, запада, западной, западные, западу, захватывает, земле, землёй, земли, землю, зупиніть , историю, киев, киева, китай, командиры, крым, лжет, лица, ложь, луганск, людей, люди, людям, майдан, мариуполе, мариуполь, мариуполя, маріуполь, мир, мире, мирном, мирные, мирными, мирных, мировая, мировой, мировую, мировые, миру, міст, народ, народа, население, населения, насилию, насилия, настоящему, наступательного, нато, невинные, невинных, невиновный, нельзя, нефти, нефть, новостей, области, обожгли, одеса, одесі, одессе, оон, остановите, остановитесь, остановить, остановка, осудила, отверстия, очнитесь, планеты, пожилой, позором, поранено, потомков, правда, правду, правительства, правительство, президент, президента, прекратите, преступлений, преступления, преступникам, преступных, проснитесь, просыпаться, родину, російський, росія, росіяни, россией, россии, российская, российские, российским, российских, российской, россию, россия, россиянам, россияне, роют, руйнують, русские, русский, русским, русских, русского, русскому, сгорели, семьи, сжечь, сжигали, слава, сми, сожгли, солдатов, специальной, ссср, стариков, стран, страна, страну, стыдно, суверенитет, суверенной, сша, территории, территорию, товарищи, угрозы, удар, ударят, фейки, фейков, харьков, харькове, херсон, человек, человечество, штурмовых, экономики, эритрея, яблунской, ядерное, ядерную} \\
      \midrule
      -3 & visitor, visitors, \foreignlanguage{russian}{посетитель, посетителям} \\
      \midrule
      -4 & beautiful, information, interesting, liked, recommend, \foreignlanguage{russian}{интересна, интересной, информации, красивое, любил, нравится, нравиться, понравилось, прекрасное, рекомендую, советую} \\
      \midrule
      -12 & attactions, attraction, exhibition, exhibits, memorial, monument, museum, tour, \foreignlanguage{russian}{достопримечательности, мемориальный, музей, памятная, памятник, памятный, привлекательность, экскурсией, экскурсию, экспозицию, экспозиция, экспонатам, экспонаты} \\
      \bottomrule
  \end{tabularx}
\end{table*}

Table~\ref{tab:classifierKeywords} lists the keywords used for the war-related content classifier and their weights for the score calculation, as described in Section~\ref{sec:contentanalysis}.

\bibliography{bibliography}

\end{document}